\documentclass[longauth]{aa}

\usepackage{graphicx}
\usepackage{natbib}
\usepackage{scalerel}

\usepackage[table]{xcolor}

\usepackage{txfonts}
\usepackage[pdfencoding=auto,psdextra]{hyperref}
\hypersetup{
    colorlinks=true,
    linkcolor=blue,
    filecolor=magenta,      
    urlcolor=blue,
    citecolor=blue
}
\makeatletter
\renewcommand*\aa@pageof{, page \thepage{} of \pageref*{LastPage}}
\makeatother

\usepackage[utf8]{inputenc}

\usepackage[switch, modulo]{lineno}
\linenumbers

\usepackage{euclid}

\newcommand*{\msun}{\si{\solarmass}}
\newcommand*{\lsun}{\si{\solarluminosity{}}}

\def\civ     {\ensuremath{\text{C\,\textsc{iv}}} }

\def\mgii     {\ensuremath{\text{Mg\,\textsc{ii}}} }

\def\cii     {\ensuremath{\text{[C\,\textsc{ii]}}}}

\def\oi    {\ensuremath{\text{[O\,\textsc{i]}}}}
\def\nii    {\ensuremath{\text{[N\,\textsc{ii]}}}}

\usepackage{booktabs}
\usepackage{threeparttable}
\usepackage{multirow}
\usepackage{physics}

\usepackage{enumitem}
\usepackage{rotating}
\usepackage{float}

\begin{document}

%
%

\title{\Euclid\/: A UV-faint quasar in a highly luminous star-forming host galaxy at $z \approx 7.7$\thanks{This paper is published on
       behalf of the Euclid Consortium}}
   

\newcommand{\orcid}[1]{} 
\author{S.~Belladitta\orcid{0000-0003-4747-4484}\thanks{\email{belladitta@mpia.de}}\inst{\ref{aff1},\ref{aff2}}
\and R.~Decarli\orcid{0000-0002-2662-8803}\inst{\ref{aff2}}
\and E.~Ba\~nados\orcid{0000-0002-2931-7824}\inst{\ref{aff1}}
\and F.~Walter\orcid{0000-0003-4793-7880}\inst{\ref{aff1}}
\and D.~Yang\orcid{0000-0002-6769-0910}\inst{\ref{aff3}}
\and F.~Guarneri\orcid{0000-0003-4740-9762}\inst{\ref{aff4},\ref{aff5}}
\and K.~Jahnke\orcid{0000-0003-3804-2137}\inst{\ref{aff1}}
\and S.~Bisogni\orcid{0000-0003-3746-4565}\inst{\ref{aff6}}
\and S.~E.~I.~Bosman\orcid{0000-0001-8582-7012}\inst{\ref{aff7},\ref{aff1}}
\and X.~Fan\orcid{0000-0003-3310-0131}\inst{\ref{aff8}}
\and Y.~Fu\orcid{0000-0002-0759-0504}\inst{\ref{aff3},\ref{aff9}}
\and J.~F.~Hennawi\orcid{0000-0002-7054-4332}\inst{\ref{aff3},\ref{aff10}}
\and Y.~Matsuoka\orcid{0000-0001-5063-0340}\inst{\ref{aff11}}
\and D.~J.~Mortlock\orcid{0000-0002-0041-3783}\inst{\ref{aff12},\ref{aff13}}
\and M.~Onoue\orcid{0000-0003-2984-6803}\inst{\ref{aff14},\ref{aff15}}
\and J.-T.~Schindler\orcid{0000-0002-4544-8242}\inst{\ref{aff4}}
\and L.~Spinoglio\orcid{0000-0001-8840-1551}\inst{\ref{aff16}}
\and D.~Stern\orcid{0000-0003-2686-9241}\inst{\ref{aff17}}
\and F.~Wang\orcid{0000-0002-7633-431X}\inst{\ref{aff18}}
\and G.~Vietri\orcid{0000-0001-9155-8875}\inst{\ref{aff6}}
\and C.~J.~Willott\orcid{0000-0002-4201-7367}\inst{\ref{aff19}}
\and J.~Wolf\orcid{0000-0003-0643-7935}\inst{\ref{aff1}}
\and J.~Yang\orcid{0000-0001-5287-4242}\inst{\ref{aff18}}
\and R.~A.~A.~Bowler\orcid{0000-0003-3917-1678}\inst{\ref{aff20}}
\and K.~I.~Caputi\orcid{0000-0001-8183-1460}\inst{\ref{aff9},\ref{aff21}}
\and D.~L.~Clements\orcid{0000-0002-9548-5033}\inst{\ref{aff12}}
\and C.~M.~Gutierrez\orcid{0000-0001-7854-783X}\inst{\ref{aff22},\ref{aff23}}
\and H.~J.~A.~Rottgering\orcid{0000-0001-8887-2257}\inst{\ref{aff3}}
\and D.~Scott\orcid{0000-0002-6878-9840}\inst{\ref{aff24}}
\and F.~Shankar\orcid{0000-0001-8973-5051}\inst{\ref{aff25}}
\and G.~Zamorani\orcid{0000-0002-2318-301X}\inst{\ref{aff2}}
\and A.-C.~Eilers\orcid{0000-0003-2895-6218}\inst{\ref{aff26},\ref{aff27}}
\and B.~Altieri\orcid{0000-0003-3936-0284}\inst{\ref{aff28}}
\and S.~Andreon\orcid{0000-0002-2041-8784}\inst{\ref{aff29}}
\and N.~Auricchio\orcid{0000-0003-4444-8651}\inst{\ref{aff2}}
\and C.~Baccigalupi\orcid{0000-0002-8211-1630}\inst{\ref{aff30},\ref{aff5},\ref{aff31},\ref{aff32}}
\and M.~Baldi\orcid{0000-0003-4145-1943}\inst{\ref{aff33},\ref{aff2},\ref{aff34}}
\and A.~Balestra\orcid{0000-0002-6967-261X}\inst{\ref{aff35}}
\and S.~Bardelli\orcid{0000-0002-8900-0298}\inst{\ref{aff2}}
\and P.~Battaglia\orcid{0000-0002-7337-5909}\inst{\ref{aff2}}
\and A.~Biviano\orcid{0000-0002-0857-0732}\inst{\ref{aff5},\ref{aff30}}
\and M.~Brescia\orcid{0000-0001-9506-5680}\inst{\ref{aff36},\ref{aff37}}
\and S.~Camera\orcid{0000-0003-3399-3574}\inst{\ref{aff38},\ref{aff39},\ref{aff40}}
\and V.~Capobianco\orcid{0000-0002-3309-7692}\inst{\ref{aff40}}
\and C.~Carbone\orcid{0000-0003-0125-3563}\inst{\ref{aff6}}
\and J.~Carretero\orcid{0000-0002-3130-0204}\inst{\ref{aff41},\ref{aff42}}
\and S.~Casas\orcid{0000-0002-4751-5138}\inst{\ref{aff43},\ref{aff44}}
\and M.~Castellano\orcid{0000-0001-9875-8263}\inst{\ref{aff45}}
\and G.~Castignani\orcid{0000-0001-6831-0687}\inst{\ref{aff2}}
\and S.~Cavuoti\orcid{0000-0002-3787-4196}\inst{\ref{aff37},\ref{aff46}}
\and K.~C.~Chambers\orcid{0000-0001-6965-7789}\inst{\ref{aff47}}
\and A.~Cimatti\inst{\ref{aff48}}
\and C.~Colodro-Conde\inst{\ref{aff22}}
\and G.~Congedo\orcid{0000-0003-2508-0046}\inst{\ref{aff49}}
\and C.~J.~Conselice\orcid{0000-0003-1949-7638}\inst{\ref{aff20}}
\and L.~Conversi\orcid{0000-0002-6710-8476}\inst{\ref{aff50},\ref{aff28}}
\and Y.~Copin\orcid{0000-0002-5317-7518}\inst{\ref{aff51}}
\and A.~Costille\inst{\ref{aff52}}
\and F.~Courbin\orcid{0000-0003-0758-6510}\inst{\ref{aff53},\ref{aff54},\ref{aff55}}
\and H.~M.~Courtois\orcid{0000-0003-0509-1776}\inst{\ref{aff56}}
\and J.-G.~Cuby\orcid{0000-0002-8767-1442}\inst{\ref{aff57},\ref{aff52}}
\and A.~Da~Silva\orcid{0000-0002-6385-1609}\inst{\ref{aff58},\ref{aff59}}
\and H.~Degaudenzi\orcid{0000-0002-5887-6799}\inst{\ref{aff60}}
\and G.~De~Lucia\orcid{0000-0002-6220-9104}\inst{\ref{aff5}}
\and H.~Dole\orcid{0000-0002-9767-3839}\inst{\ref{aff61}}
\and M.~Douspis\orcid{0000-0003-4203-3954}\inst{\ref{aff61}}
\and F.~Dubath\orcid{0000-0002-6533-2810}\inst{\ref{aff60}}
\and X.~Dupac\inst{\ref{aff28}}
\and S.~Dusini\orcid{0000-0002-1128-0664}\inst{\ref{aff62}}
\and S.~Escoffier\orcid{0000-0002-2847-7498}\inst{\ref{aff63}}
\and M.~Farina\orcid{0000-0002-3089-7846}\inst{\ref{aff16}}
\and R.~Farinelli\inst{\ref{aff2}}
\and F.~Faustini\orcid{0000-0001-6274-5145}\inst{\ref{aff45},\ref{aff64}}
\and S.~Ferriol\inst{\ref{aff51}}
\and F.~Finelli\orcid{0000-0002-6694-3269}\inst{\ref{aff2},\ref{aff65}}
\and P.~Fosalba\orcid{0000-0002-1510-5214}\inst{\ref{aff66},\ref{aff67}}
\and S.~Fotopoulou\orcid{0000-0002-9686-254X}\inst{\ref{aff68}}
\and M.~Frailis\orcid{0000-0002-7400-2135}\inst{\ref{aff5}}
\and E.~Franceschi\orcid{0000-0002-0585-6591}\inst{\ref{aff2}}
\and M.~Fumana\orcid{0000-0001-6787-5950}\inst{\ref{aff6}}
\and S.~Galeotta\orcid{0000-0002-3748-5115}\inst{\ref{aff5}}
\and K.~George\orcid{0000-0002-1734-8455}\inst{\ref{aff69}}
\and B.~Gillis\orcid{0000-0002-4478-1270}\inst{\ref{aff49}}
\and C.~Giocoli\orcid{0000-0002-9590-7961}\inst{\ref{aff2},\ref{aff34}}
\and J.~Gracia-Carpio\orcid{0000-0003-4689-3134}\inst{\ref{aff70}}
\and A.~Grazian\orcid{0000-0002-5688-0663}\inst{\ref{aff35}}
\and F.~Grupp\inst{\ref{aff70},\ref{aff71}}
\and L.~Guzzo\orcid{0000-0001-8264-5192}\inst{\ref{aff72},\ref{aff29},\ref{aff73}}
\and S.~V.~H.~Haugan\orcid{0000-0001-9648-7260}\inst{\ref{aff74}}
\and H.~Hoekstra\orcid{0000-0002-0641-3231}\inst{\ref{aff3}}
\and W.~Holmes\inst{\ref{aff17}}
\and I.~M.~Hook\orcid{0000-0002-2960-978X}\inst{\ref{aff75}}
\and F.~Hormuth\inst{\ref{aff76}}
\and A.~Hornstrup\orcid{0000-0002-3363-0936}\inst{\ref{aff77},\ref{aff78}}
\and M.~Jhabvala\inst{\ref{aff79}}
\and B.~Joachimi\orcid{0000-0001-7494-1303}\inst{\ref{aff80}}
\and S.~Kermiche\orcid{0000-0002-0302-5735}\inst{\ref{aff63}}
\and A.~Kiessling\orcid{0000-0002-2590-1273}\inst{\ref{aff17}}
\and B.~Kubik\orcid{0009-0006-5823-4880}\inst{\ref{aff51}}
\and M.~K\"ummel\orcid{0000-0003-2791-2117}\inst{\ref{aff71}}
\and M.~Kunz\orcid{0000-0002-3052-7394}\inst{\ref{aff81}}
\and H.~Kurki-Suonio\orcid{0000-0002-4618-3063}\inst{\ref{aff82},\ref{aff83}}
\and R.~Laureijs\inst{\ref{aff9}}
\and A.~M.~C.~Le~Brun\orcid{0000-0002-0936-4594}\inst{\ref{aff84}}
\and S.~Ligori\orcid{0000-0003-4172-4606}\inst{\ref{aff40}}
\and P.~B.~Lilje\orcid{0000-0003-4324-7794}\inst{\ref{aff74}}
\and V.~Lindholm\orcid{0000-0003-2317-5471}\inst{\ref{aff82},\ref{aff83}}
\and I.~Lloro\orcid{0000-0001-5966-1434}\inst{\ref{aff85}}
\and G.~Mainetti\orcid{0000-0003-2384-2377}\inst{\ref{aff86}}
\and D.~Maino\inst{\ref{aff72},\ref{aff6},\ref{aff73}}
\and E.~Maiorano\orcid{0000-0003-2593-4355}\inst{\ref{aff2}}
\and O.~Mansutti\orcid{0000-0001-5758-4658}\inst{\ref{aff5}}
\and S.~Marcin\inst{\ref{aff87}}
\and O.~Marggraf\orcid{0000-0001-7242-3852}\inst{\ref{aff88}}
\and M.~Martinelli\orcid{0000-0002-6943-7732}\inst{\ref{aff45},\ref{aff89}}
\and N.~Martinet\orcid{0000-0003-2786-7790}\inst{\ref{aff52}}
\and F.~Marulli\orcid{0000-0002-8850-0303}\inst{\ref{aff90},\ref{aff2},\ref{aff34}}
\and R.~J.~Massey\orcid{0000-0002-6085-3780}\inst{\ref{aff91}}
\and E.~Medinaceli\orcid{0000-0002-4040-7783}\inst{\ref{aff2}}
\and S.~Mei\orcid{0000-0002-2849-559X}\inst{\ref{aff92},\ref{aff93}}
\and M.~Melchior\inst{\ref{aff94}}
\and M.~Meneghetti\orcid{0000-0003-1225-7084}\inst{\ref{aff2},\ref{aff34}}
\and E.~Merlin\orcid{0000-0001-6870-8900}\inst{\ref{aff45}}
\and G.~Meylan\inst{\ref{aff95}}
\and A.~Mora\orcid{0000-0002-1922-8529}\inst{\ref{aff96}}
\and M.~Moresco\orcid{0000-0002-7616-7136}\inst{\ref{aff90},\ref{aff2}}
\and L.~Moscardini\orcid{0000-0002-3473-6716}\inst{\ref{aff90},\ref{aff2},\ref{aff34}}
\and C.~Neissner\orcid{0000-0001-8524-4968}\inst{\ref{aff97},\ref{aff42}}
\and R.~C.~Nichol\orcid{0000-0003-0939-6518}\inst{\ref{aff98}}
\and S.-M.~Niemi\orcid{0009-0005-0247-0086}\inst{\ref{aff99}}
\and C.~Padilla\orcid{0000-0001-7951-0166}\inst{\ref{aff97}}
\and S.~Paltani\orcid{0000-0002-8108-9179}\inst{\ref{aff60}}
\and F.~Pasian\orcid{0000-0002-4869-3227}\inst{\ref{aff5}}
\and K.~Pedersen\inst{\ref{aff100}}
\and W.~J.~Percival\orcid{0000-0002-0644-5727}\inst{\ref{aff101},\ref{aff102},\ref{aff103}}
\and V.~Pettorino\orcid{0000-0002-4203-9320}\inst{\ref{aff99}}
\and S.~Pires\orcid{0000-0002-0249-2104}\inst{\ref{aff104}}
\and G.~Polenta\orcid{0000-0003-4067-9196}\inst{\ref{aff64}}
\and M.~Poncet\inst{\ref{aff105}}
\and L.~A.~Popa\inst{\ref{aff106}}
\and L.~Pozzetti\orcid{0000-0001-7085-0412}\inst{\ref{aff2}}
\and F.~Raison\orcid{0000-0002-7819-6918}\inst{\ref{aff70}}
\and A.~Renzi\orcid{0000-0001-9856-1970}\inst{\ref{aff107},\ref{aff62}}
\and J.~Rhodes\orcid{0000-0002-4485-8549}\inst{\ref{aff17}}
\and G.~Riccio\inst{\ref{aff37}}
\and H.-W.~Rix\orcid{0000-0003-4996-9069}\inst{\ref{aff1}}
\and E.~Romelli\orcid{0000-0003-3069-9222}\inst{\ref{aff5}}
\and M.~Roncarelli\orcid{0000-0001-9587-7822}\inst{\ref{aff2}}
\and B.~Rusholme\orcid{0000-0001-7648-4142}\inst{\ref{aff108}}
\and R.~Saglia\orcid{0000-0003-0378-7032}\inst{\ref{aff71},\ref{aff70}}
\and Z.~Sakr\orcid{0000-0002-4823-3757}\inst{\ref{aff7},\ref{aff109},\ref{aff110}}
\and D.~Sapone\orcid{0000-0001-7089-4503}\inst{\ref{aff111}}
\and B.~Sartoris\orcid{0000-0003-1337-5269}\inst{\ref{aff71},\ref{aff5}}
\and M.~Schirmer\orcid{0000-0003-2568-9994}\inst{\ref{aff1}}
\and P.~Schneider\orcid{0000-0001-8561-2679}\inst{\ref{aff88}}
\and T.~Schrabback\orcid{0000-0002-6987-7834}\inst{\ref{aff112}}
\and A.~Secroun\orcid{0000-0003-0505-3710}\inst{\ref{aff63}}
\and G.~Seidel\orcid{0000-0003-2907-353X}\inst{\ref{aff1}}
\and E.~Sihvola\orcid{0000-0003-1804-7715}\inst{\ref{aff113}}
\and P.~Simon\inst{\ref{aff88}}
\and C.~Sirignano\orcid{0000-0002-0995-7146}\inst{\ref{aff107},\ref{aff62}}
\and G.~Sirri\orcid{0000-0003-2626-2853}\inst{\ref{aff34}}
\and L.~Stanco\orcid{0000-0002-9706-5104}\inst{\ref{aff62}}
\and P.~Tallada-Cresp\'{i}\orcid{0000-0002-1336-8328}\inst{\ref{aff41},\ref{aff42}}
\and A.~N.~Taylor\inst{\ref{aff49}}
\and I.~Tereno\orcid{0000-0002-4537-6218}\inst{\ref{aff58},\ref{aff114}}
\and N.~Tessore\orcid{0000-0002-9696-7931}\inst{\ref{aff115}}
\and S.~Toft\orcid{0000-0003-3631-7176}\inst{\ref{aff21},\ref{aff116}}
\and R.~Toledo-Moreo\orcid{0000-0002-2997-4859}\inst{\ref{aff117}}
\and F.~Torradeflot\orcid{0000-0003-1160-1517}\inst{\ref{aff42},\ref{aff41}}
\and I.~Tutusaus\orcid{0000-0002-3199-0399}\inst{\ref{aff67},\ref{aff66},\ref{aff109}}
\and L.~Valenziano\orcid{0000-0002-1170-0104}\inst{\ref{aff2},\ref{aff65}}
\and J.~Valiviita\orcid{0000-0001-6225-3693}\inst{\ref{aff82},\ref{aff83}}
\and T.~Vassallo\orcid{0000-0001-6512-6358}\inst{\ref{aff5},\ref{aff69}}
\and Y.~Wang\orcid{0000-0002-4749-2984}\inst{\ref{aff108}}
\and J.~Weller\orcid{0000-0002-8282-2010}\inst{\ref{aff71},\ref{aff70}}
\and F.~M.~Zerbi\orcid{0000-0002-9996-973X}\inst{\ref{aff29}}
\and E.~Zucca\orcid{0000-0002-5845-8132}\inst{\ref{aff2}}
\and J.~Garc\'ia-Bellido\orcid{0000-0002-9370-8360}\inst{\ref{aff118}}
\and J.~Mart\'{i}n-Fleitas\orcid{0000-0002-8594-569X}\inst{\ref{aff119}}
\and P.~Monaco\orcid{0000-0003-2083-7564}\inst{\ref{aff120},\ref{aff5},\ref{aff31},\ref{aff30}}
\and V.~Scottez\orcid{0009-0008-3864-940X}\inst{\ref{aff121},\ref{aff122}}
\and M.~Viel\orcid{0000-0002-2642-5707}\inst{\ref{aff30},\ref{aff5},\ref{aff32},\ref{aff31},\ref{aff123}}}
                                                                                   
\institute{Max-Planck-Institut f\"ur Astronomie, K\"onigstuhl 17, 69117 Heidelberg, Germany\label{aff1}
\and
INAF-Osservatorio di Astrofisica e Scienza dello Spazio di Bologna, Via Piero Gobetti 93/3, 40129 Bologna, Italy\label{aff2}
\and
Leiden Observatory, Leiden University, Einsteinweg 55, 2333 CC Leiden, The Netherlands\label{aff3}
\and
Hamburger Sternwarte, University of Hamburg, Gojenbergsweg 112, 21029 Hamburg, Germany\label{aff4}
\and
INAF-Osservatorio Astronomico di Trieste, Via G. B. Tiepolo 11, 34143 Trieste, Italy\label{aff5}
\and
INAF-IASF Milano, Via Alfonso Corti 12, 20133 Milano, Italy\label{aff6}
\and
Institut f\"ur Theoretische Physik, University of Heidelberg, Philosophenweg 16, 69120 Heidelberg, Germany\label{aff7}
\and
Steward Observatory, University of Arizona, 933 N. Cherry Ave, Tucson, AZ 85750, USA\label{aff8}
\and
Kapteyn Astronomical Institute, University of Groningen, PO Box 800, 9700 AV Groningen, The Netherlands\label{aff9}
\and
Department of Physics, University of California, Santa Barbara, CA 93106, USA\label{aff10}
\and
Research Center for Space and Cosmic Evolution, Ehime University, 2-5 Bunkyo-cho, Matsuyama, Ehime 790-8577, Japan\label{aff11}
\and
Astrophysics Group, Blackett Laboratory, Imperial College London, London SW7 2AZ, UK\label{aff12}
\and
Department of Mathematics, Imperial College London, London SW7 2AZ, UK\label{aff13}
\and
Waseda Institute for Advanced Study (WIAS), Waseda University, 1-21-1, Nishi-Waseda, Shinjuku, Tokyo 169-0051, Japan\label{aff14}
\and
Kavli Institute for the Physics and Mathematics of the Universe (WPI), University of Tokyo, Kashiwa, Chiba 277-8583, Japan\label{aff15}
\and
INAF-Istituto di Astrofisica e Planetologia Spaziali, via del Fosso del Cavaliere, 100, 00100 Roma, Italy\label{aff16}
\and
Jet Propulsion Laboratory, California Institute of Technology, 4800 Oak Grove Drive, Pasadena, CA, 91109, USA\label{aff17}
\and
Department of Astronomy, University of Michigan, 1085 S. University Ave., Ann Arbor, MI 48109, USA\label{aff18}
\and
Herzberg Astronomy and Astrophysics Research Centre, 5071 W. Saanich Rd. Victoria, BC, V9E 2E7, Canada\label{aff19}
\and
Jodrell Bank Centre for Astrophysics, Department of Physics and Astronomy, University of Manchester, Oxford Road, Manchester M13 9PL, UK\label{aff20}
\and
Cosmic Dawn Center (DAWN)\label{aff21}
\and
Instituto de Astrof\'{\i}sica de Canarias, E-38205 La Laguna, Tenerife, Spain\label{aff22}
\and
Universidad de La Laguna, Dpto. Astrof\'\i sica, E-38206 La Laguna, Tenerife, Spain\label{aff23}
\and
Department of Physics and Astronomy, University of British Columbia, Vancouver, BC V6T 1Z1, Canada\label{aff24}
\and
School of Physics \& Astronomy, University of Southampton, Highfield Campus, Southampton SO17 1BJ, UK\label{aff25}
\and
Department of Physics, Massachusetts Institute of Technology, Cambridge, MA 02139, USA\label{aff26}
\and
MIT Kavli Institute for Astrophysics and Space Research, Massachusetts Institute of Technology, Cambridge, MA 02139, USA\label{aff27}
\and
ESAC/ESA, Camino Bajo del Castillo, s/n., Urb. Villafranca del Castillo, 28692 Villanueva de la Ca\~nada, Madrid, Spain\label{aff28}
\and
INAF-Osservatorio Astronomico di Brera, Via Brera 28, 20122 Milano, Italy\label{aff29}
\and
IFPU, Institute for Fundamental Physics of the Universe, via Beirut 2, 34151 Trieste, Italy\label{aff30}
\and
INFN, Sezione di Trieste, Via Valerio 2, 34127 Trieste TS, Italy\label{aff31}
\and
SISSA, International School for Advanced Studies, Via Bonomea 265, 34136 Trieste TS, Italy\label{aff32}
\and
Dipartimento di Fisica e Astronomia, Universit\`a di Bologna, Via Gobetti 93/2, 40129 Bologna, Italy\label{aff33}
\and
INFN-Sezione di Bologna, Viale Berti Pichat 6/2, 40127 Bologna, Italy\label{aff34}
\and
INAF-Osservatorio Astronomico di Padova, Via dell'Osservatorio 5, 35122 Padova, Italy\label{aff35}
\and
Department of Physics "E. Pancini", University Federico II, Via Cinthia 6, 80126, Napoli, Italy\label{aff36}
\and
INAF-Osservatorio Astronomico di Capodimonte, Via Moiariello 16, 80131 Napoli, Italy\label{aff37}
\and
Dipartimento di Fisica, Universit\`a degli Studi di Torino, Via P. Giuria 1, 10125 Torino, Italy\label{aff38}
\and
INFN-Sezione di Torino, Via P. Giuria 1, 10125 Torino, Italy\label{aff39}
\and
INAF-Osservatorio Astrofisico di Torino, Via Osservatorio 20, 10025 Pino Torinese (TO), Italy\label{aff40}
\and
Centro de Investigaciones Energ\'eticas, Medioambientales y Tecnol\'ogicas (CIEMAT), Avenida Complutense 40, 28040 Madrid, Spain\label{aff41}
\and
Port d'Informaci\'{o} Cient\'{i}fica, Campus UAB, C. Albareda s/n, 08193 Bellaterra (Barcelona), Spain\label{aff42}
\and
Institute for Theoretical Particle Physics and Cosmology (TTK), RWTH Aachen University, 52056 Aachen, Germany\label{aff43}
\and
Deutsches Zentrum f\"ur Luft- und Raumfahrt e. V. (DLR), Linder H\"ohe, 51147 K\"oln, Germany\label{aff44}
\and
INAF-Osservatorio Astronomico di Roma, Via Frascati 33, 00078 Monteporzio Catone, Italy\label{aff45}
\and
INFN section of Naples, Via Cinthia 6, 80126, Napoli, Italy\label{aff46}
\and
Institute for Astronomy, University of Hawaii, 2680 Woodlawn Drive, Honolulu, HI 96822, USA\label{aff47}
\and
Dipartimento di Fisica e Astronomia "Augusto Righi" - Alma Mater Studiorum Universit\`a di Bologna, Viale Berti Pichat 6/2, 40127 Bologna, Italy\label{aff48}
\and
Institute for Astronomy, University of Edinburgh, Royal Observatory, Blackford Hill, Edinburgh EH9 3HJ, UK\label{aff49}
\and
European Space Agency/ESRIN, Largo Galileo Galilei 1, 00044 Frascati, Roma, Italy\label{aff50}
\and
Universit\'e Claude Bernard Lyon 1, CNRS/IN2P3, IP2I Lyon, UMR 5822, Villeurbanne, F-69100, France\label{aff51}
\and
Aix-Marseille Universit\'e, CNRS, CNES, LAM, Marseille, France\label{aff52}
\and
Institut de Ci\`{e}ncies del Cosmos (ICCUB), Universitat de Barcelona (IEEC-UB), Mart\'{i} i Franqu\`{e}s 1, 08028 Barcelona, Spain\label{aff53}
\and
Instituci\'o Catalana de Recerca i Estudis Avan\c{c}ats (ICREA), Passeig de Llu\'{\i}s Companys 23, 08010 Barcelona, Spain\label{aff54}
\and
Institut de Ciencies de l'Espai (IEEC-CSIC), Campus UAB, Carrer de Can Magrans, s/n Cerdanyola del Vall\'es, 08193 Barcelona, Spain\label{aff55}
\and
UCB Lyon 1, CNRS/IN2P3, IUF, IP2I Lyon, 4 rue Enrico Fermi, 69622 Villeurbanne, France\label{aff56}
\and
Canada-France-Hawaii Telescope, 65-1238 Mamalahoa Hwy, Kamuela, HI 96743, USA\label{aff57}
\and
Departamento de F\'isica, Faculdade de Ci\^encias, Universidade de Lisboa, Edif\'icio C8, Campo Grande, PT1749-016 Lisboa, Portugal\label{aff58}
\and
Instituto de Astrof\'isica e Ci\^encias do Espa\c{c}o, Faculdade de Ci\^encias, Universidade de Lisboa, Campo Grande, 1749-016 Lisboa, Portugal\label{aff59}
\and
Department of Astronomy, University of Geneva, ch. d'Ecogia 16, 1290 Versoix, Switzerland\label{aff60}
\and
Universit\'e Paris-Saclay, CNRS, Institut d'astrophysique spatiale, 91405, Orsay, France\label{aff61}
\and
INFN-Padova, Via Marzolo 8, 35131 Padova, Italy\label{aff62}
\and
Aix-Marseille Universit\'e, CNRS/IN2P3, CPPM, Marseille, France\label{aff63}
\and
Space Science Data Center, Italian Space Agency, via del Politecnico snc, 00133 Roma, Italy\label{aff64}
\and
INFN-Bologna, Via Irnerio 46, 40126 Bologna, Italy\label{aff65}
\and
Institut d'Estudis Espacials de Catalunya (IEEC),  Edifici RDIT, Campus UPC, 08860 Castelldefels, Barcelona, Spain\label{aff66}
\and
Institute of Space Sciences (ICE, CSIC), Campus UAB, Carrer de Can Magrans, s/n, 08193 Barcelona, Spain\label{aff67}
\and
School of Physics, HH Wills Physics Laboratory, University of Bristol, Tyndall Avenue, Bristol, BS8 1TL, UK\label{aff68}
\and
University Observatory, LMU Faculty of Physics, Scheinerstr.~1, 81679 Munich, Germany\label{aff69}
\and
Max Planck Institute for Extraterrestrial Physics, Giessenbachstr. 1, 85748 Garching, Germany\label{aff70}
\and
Universit\"ats-Sternwarte M\"unchen, Fakult\"at f\"ur Physik, Ludwig-Maximilians-Universit\"at M\"unchen, Scheinerstr.~1, 81679 M\"unchen, Germany\label{aff71}
\and
Dipartimento di Fisica "Aldo Pontremoli", Universit\`a degli Studi di Milano, Via Celoria 16, 20133 Milano, Italy\label{aff72}
\and
INFN-Sezione di Milano, Via Celoria 16, 20133 Milano, Italy\label{aff73}
\and
Institute of Theoretical Astrophysics, University of Oslo, P.O. Box 1029 Blindern, 0315 Oslo, Norway\label{aff74}
\and
Department of Physics, Lancaster University, Lancaster, LA1 4YB, UK\label{aff75}
\and
Felix Hormuth Engineering, Goethestr. 17, 69181 Leimen, Germany\label{aff76}
\and
Technical University of Denmark, Elektrovej 327, 2800 Kgs. Lyngby, Denmark\label{aff77}
\and
Cosmic Dawn Center (DAWN), Denmark\label{aff78}
\and
NASA Goddard Space Flight Center, Greenbelt, MD 20771, USA\label{aff79}
\and
Department of Physics and Astronomy, University College London, Gower Street, London WC1E 6BT, UK\label{aff80}
\and
Universit\'e de Gen\`eve, D\'epartement de Physique Th\'eorique and Centre for Astroparticle Physics, 24 quai Ernest-Ansermet, CH-1211 Gen\`eve 4, Switzerland\label{aff81}
\and
Department of Physics, P.O. Box 64, University of Helsinki, 00014 Helsinki, Finland\label{aff82}
\and
Helsinki Institute of Physics, Gustaf H{\"a}llstr{\"o}min katu 2, University of Helsinki, 00014 Helsinki, Finland\label{aff83}
\and
Laboratoire d'etude de l'Univers et des phenomenes eXtremes, Observatoire de Paris, Universit\'e PSL, Sorbonne Universit\'e, CNRS, 92190 Meudon, France\label{aff84}
\and
SKAO, Jodrell Bank, Lower Withington, Macclesfield SK11 9FT, UK\label{aff85}
\and
Centre de Calcul de l'IN2P3/CNRS, 21 avenue Pierre de Coubertin 69627 Villeurbanne Cedex, France\label{aff86}
\and
University of Applied Sciences and Arts of Northwestern Switzerland, School of Computer Science, 5210 Windisch, Switzerland\label{aff87}
\and
Universit\"at Bonn, Argelander-Institut f\"ur Astronomie, Auf dem H\"ugel 71, 53121 Bonn, Germany\label{aff88}
\and
INFN-Sezione di Roma, Piazzale Aldo Moro, 2 - c/o Dipartimento di Fisica, Edificio G. Marconi, 00185 Roma, Italy\label{aff89}
\and
Dipartimento di Fisica e Astronomia "Augusto Righi" - Alma Mater Studiorum Universit\`a di Bologna, via Piero Gobetti 93/2, 40129 Bologna, Italy\label{aff90}
\and
Department of Physics, Institute for Computational Cosmology, Durham University, South Road, Durham, DH1 3LE, UK\label{aff91}
\and
Universit\'e Paris Cit\'e, CNRS, Astroparticule et Cosmologie, 75013 Paris, France\label{aff92}
\and
CNRS-UCB International Research Laboratory, Centre Pierre Bin\'etruy, IRL2007, CPB-IN2P3, Berkeley, USA\label{aff93}
\and
University of Applied Sciences and Arts of Northwestern Switzerland, School of Engineering, 5210 Windisch, Switzerland\label{aff94}
\and
Institute of Physics, Laboratory of Astrophysics, Ecole Polytechnique F\'ed\'erale de Lausanne (EPFL), Observatoire de Sauverny, 1290 Versoix, Switzerland\label{aff95}
\and
Telespazio UK S.L. for European Space Agency (ESA), Camino bajo del Castillo, s/n, Urbanizacion Villafranca del Castillo, Villanueva de la Ca\~nada, 28692 Madrid, Spain\label{aff96}
\and
Institut de F\'{i}sica d'Altes Energies (IFAE), The Barcelona Institute of Science and Technology, Campus UAB, 08193 Bellaterra (Barcelona), Spain\label{aff97}
\and
School of Mathematics and Physics, University of Surrey, Guildford, Surrey, GU2 7XH, UK\label{aff98}
\and
European Space Agency/ESTEC, Keplerlaan 1, 2201 AZ Noordwijk, The Netherlands\label{aff99}
\and
DARK, Niels Bohr Institute, University of Copenhagen, Jagtvej 155, 2200 Copenhagen, Denmark\label{aff100}
\and
Waterloo Centre for Astrophysics, University of Waterloo, Waterloo, Ontario N2L 3G1, Canada\label{aff101}
\and
Department of Physics and Astronomy, University of Waterloo, Waterloo, Ontario N2L 3G1, Canada\label{aff102}
\and
Perimeter Institute for Theoretical Physics, Waterloo, Ontario N2L 2Y5, Canada\label{aff103}
\and
Universit\'e Paris-Saclay, Universit\'e Paris Cit\'e, CEA, CNRS, AIM, 91191, Gif-sur-Yvette, France\label{aff104}
\and
Centre National d'Etudes Spatiales -- Centre spatial de Toulouse, 18 avenue Edouard Belin, 31401 Toulouse Cedex 9, France\label{aff105}
\and
Institute of Space Science, Str. Atomistilor, nr. 409 M\u{a}gurele, Ilfov, 077125, Romania\label{aff106}
\and
Dipartimento di Fisica e Astronomia "G. Galilei", Universit\`a di Padova, Via Marzolo 8, 35131 Padova, Italy\label{aff107}
\and
Caltech/IPAC, 1200 E. California Blvd., Pasadena, CA 91125, USA\label{aff108}
\and
Institut de Recherche en Astrophysique et Plan\'etologie (IRAP), Universit\'e de Toulouse, CNRS, UPS, CNES, 14 Av. Edouard Belin, 31400 Toulouse, France\label{aff109}
\and
Universit\'e St Joseph; Faculty of Sciences, Beirut, Lebanon\label{aff110}
\and
Departamento de F\'isica, FCFM, Universidad de Chile, Blanco Encalada 2008, Santiago, Chile\label{aff111}
\and
Universit\"at Innsbruck, Institut f\"ur Astro- und Teilchenphysik, Technikerstr. 25/8, 6020 Innsbruck, Austria\label{aff112}
\and
Department of Physics and Helsinki Institute of Physics, Gustaf H\"allstr\"omin katu 2, University of Helsinki, 00014 Helsinki, Finland\label{aff113}
\and
Instituto de Astrof\'isica e Ci\^encias do Espa\c{c}o, Faculdade de Ci\^encias, Universidade de Lisboa, Tapada da Ajuda, 1349-018 Lisboa, Portugal\label{aff114}
\and
Mullard Space Science Laboratory, University College London, Holmbury St Mary, Dorking, Surrey RH5 6NT, UK\label{aff115}
\and
Niels Bohr Institute, University of Copenhagen, Jagtvej 128, 2200 Copenhagen, Denmark\label{aff116}
\and
Universidad Polit\'ecnica de Cartagena, Departamento de Electr\'onica y Tecnolog\'ia de Computadoras,  Plaza del Hospital 1, 30202 Cartagena, Spain\label{aff117}
\and
Instituto de F\'isica Te\'orica UAM-CSIC, Campus de Cantoblanco, 28049 Madrid, Spain\label{aff118}
\and
Aurora Technology for European Space Agency (ESA), Camino bajo del Castillo, s/n, Urbanizacion Villafranca del Castillo, Villanueva de la Ca\~nada, 28692 Madrid, Spain\label{aff119}
\and
Dipartimento di Fisica - Sezione di Astronomia, Universit\`a di Trieste, Via Tiepolo 11, 34131 Trieste, Italy\label{aff120}
\and
Institut d'Astrophysique de Paris, 98bis Boulevard Arago, 75014, Paris, France\label{aff121}
\and
ICL, Junia, Universit\'e Catholique de Lille, LITL, 59000 Lille, France\label{aff122}
\and
ICSC - Centro Nazionale di Ricerca in High Performance Computing, Big Data e Quantum Computing, Via Magnanelli 2, Bologna, Italy\label{aff123}}    

%
%
\abstract{Constraining the co-evolution of supermassive black holes and their host galaxies in the first billion years after the Big Bang is essential for understanding the formation of the earliest cosmic structures. Here, we present IRAM/NOrthern Extended Millimeter Array (NOEMA) observations of the $z \approx 7.7$ quasar EUCL\,J125308.55+705432.3, recently discovered in the first data release of the Euclid Wide Survey. We report the most distant detections of \cii\ 158\,\micron\ and cold dust emission in a quasar host to date. The \cii\ emission line sets the systemic redshift at $z=7.6980\pm0.0004$. The source exhibits luminosities of $L_{\rm FIR}=3.6\times10^{12}\,L_{\odot}$ and $L_{\cii}=2\times10^9\,L_{\odot}$, respectively, a dust mass of $1.4\times 10^{8} \msun$, and a dynamical mass in the range $0.33-1.3\times10^{10} \msun$.
Remarkably, despite being nearly two magnitudes fainter in the rest-frame UV ($M_{1450}=-24.06$) than previously known $z\approx7.5$ quasars ($\ave{M_{1450}}$ $\sim-$26.5), 
EUCL\,J125308.55+705432.3 exhibits the brightest \cii\ emission among them. This indicates that the host galaxy is actively star-forming, with a star-formation rate $>250\,M_{\sun}\,\mathrm{yr}^{-1}$, consistent with recent findings that UV-faint quasars at $z>6$ preferentially reside in \cii-luminous galaxies. 
The UV-faintness likely reflects dust obscuration or sub-Eddington accretion, rather than lower host mass, suggesting these systems are at a different stage in their evolution compared to UV-bright quasars.
These IRAM/NOEMA observations highlight the power of combining \Euclid's wide-area quasar discovery potential with submillimetre follow-up observations to characterise the host galaxies of early supermassive black holes across a broader redshift and luminosity range than previously accessible.
}

%
%
%
%
   \titlerunning{The host galaxy of a UV-faint $z=7.7$ quasar}
   \authorrunning{S.~Belladitta et al.}
   
   \maketitle
   \nolinenumbers
%
%
%
%
   
\section{\label{sec:Intro}Introduction}
The most distant quasars, residing at the epoch of reionisation (EoR, at redshift $z>6$), can serve as powerful probes of early Universe conditions. They provide essential benchmarks for models of supermassive black hole (SMBH) formation and growth, galaxy assembly, large-scale structure evolution, along with the history of cosmic reionisation and chemical enrichment of the intergalactic medium \citep[see][for a recent review]{fan2023}.
To date, only three systems at $z\approx7.5$ anchor the current quasar redshift frontier \citep{banados2018,yang2020,wang2021}.
These UV-bright ($\ave{M_{1450}}\simeq$ $-$26.5) sources harbour SMBHs with masses exceeding 10$^9$\,\msun\ \citep[e.g.][]{yang2021,bosman2025}, challenging our understanding of how such massive objects could assemble within the first 700 million years of the Universe's existence \citep[e.g.][]{inayoshi2020}.

The accreting SMBH injects large amounts of energy into the surrounding medium, influencing gas dynamics and star formation within the host galaxy and thereby regulating its dynamical properties and baryonic content \citep[e.g.][]{dimatteo2005,hopkins2008}. 
Probing the SMBH--galaxy relation at the highest redshift strongly benefit from submillimetre observations with the Atacama Large Millimeter/submillimeter Array (ALMA) and/or the NOrthern Extended Millimeter Array (NOEMA), which provide key diagnostics of the interstellar medium~(ISM). 

All three known quasars at $z\approx 7.5$ have been observed with ALMA and/or NOEMA to investigate their host galaxies \citep{venemans2017,banados2019,novak2019,yang2020,wang2021,feruglio2023,wang2024,salvestrini2025}.  
These studies, together with similar efforts on lower redshift sources \citep[e.g.][]{decarli2018,venemans2018,venemans2020,neeleman2021,izumi2021a,yue2021,banados2024}, reveal that early quasar host galaxies share several remarkable properties:
(i) they are gas-rich systems undergoing vigorous star formation, with star-formation rates (SFRs) ranging from 50 to 1000\,$ \msun\,\mathrm{yr}^{-1}$; 
(ii) they are bright in the far-infrared (FIR), with luminosities of $L_\mathrm{FIR}\approx10^{12-13}\,\lsun$, indicative of intense dust heating driven by elevated SFRs; 
(iii) the inferred dust temperatures (30--60\,K) and dust masses ($\gtrsim10^{8}\,\msun$) suggest the presence of massive molecular gas reservoirs; 
(iv) these systems exhibit strong \cii\ 158\,\micron\ emission, a principal coolant of the ISM that directly traces both star formation activity and the kinematics of atomic and ionised gas \citep[e.g.][]{diazsantos2016};
(v) their \cii/FIR luminosity ratios are consistent with those found in high-redshift starbursts and local ultraluminous infrared galaxies (ULIRGs), reflecting the combined influence of accretion-powered heating and star formation in shaping ISM conditions;
(vi) high-resolution \cii\ imaging on scales of about 200--300\,pc reveals rotating gas discs with compact, structures.  
Kinematic modelling of these high-resolution data suggests that the most luminous systems harbour overmassive SMBHs relative to expectations from the local $M$--$\sigma_{\star}$ relation; and (vii) the detection of dust-rich companion galaxies near to the central quasar points to galaxy-rich environments at early cosmic times, implying that major mergers may have played a key role in fuelling the rapid growth of both SMBHs and their host galaxies.

Despite the availability of extensive studies, the rapid co-evolution of massive galaxies and their central SMBHs remains poorly understood, particularly in terms of the properties, dynamics, and fuelling efficiency of their molecular gas.
While the three quasars discovered at $z\approx7.5$ have provided crucial insights into this epoch, this small sample limits our ability to constrain early galaxy and black hole assembly models at the redshifts where they are most discriminating.
The \Euclid mission \citep[][]{EuclidSkyOverview} is dramatically expanding this sample, having identified 31 new quasars at $6.6\leq z \leq 7.8$ in its first 1.5 year, more than doubling the number of known $z>7$ quasars \citep[][]{yang2026}.

Among these discoveries is EUCL\,J125308.55+705432.3 (hereafter EUCL\,J1253+7054) at $z\approx7.7$, the second most distant quasar known to date and the focus of this paper. 
Remarkably, EUCL\,J1253+7054 is nearly two magnitudes fainter in
the rest-frame UV ($M_{1450} = -24.06$) than the previous known $z\approx7.5$ quasars, demonstrating \Euclid's ability to probe the uncharted territory of UV-faint quasars at $z>7$. 
On the other hand EUCL\,J1253+7054 is also an order magnitude brighter in the UV than the brightest ($\ave{M_{1450}}$ $\approx-22$) high-redshift galaxies discovered so far \citep[e.g.][but see also \citealt{yang2026}]{bouwens2022,matsuoka2025,robertsborsani2024,robertsborsani2025}.

Such UV-faint ($M_{1450}\gtrsim-24$) quasars as EUCL\,J1253+7054 could represent an important link between the most luminous quasars ($M_{1450}\approx-26$) and the fainter ($M_{1450}\approx-20$) broad-line active galactic nuclei (AGNs) commonly detected by the {\it James Webb} Space Telescope (JWST) in the EoR \citep[e.g.][]{harikane2023,maiolino2024,schindler2025}. 
At this intermediate luminosity, we probe a regime where the AGN still dominates the total luminosity budget, but the central quasar is not outshining the host galaxy as severely as in the most luminous systems \citep[e.g.][]{ding2025}. 
Moreover, this magnitude range represents a transition zone where the UV luminosity functions of quasars and star-forming galaxies begin to overlap \citep[e.g.][]{matsuoka2023}, making UV-faint quasars a crucial link for understanding the evolution of both populations.

In this paper, we report the most distant \cii\ and cold dust observations of a quasar host galaxy to date, obtained with IRAM/NOEMA, and we characterise the properties of the EUCL\,J1253$+$7054 system.
For full details on the selection, discovery and optical/NIR properties of EUCL\,J1253$+$7054, we refer to \citet[][]{yang2026}.

The paper is structured as follows. In Sect.~\ref{sec:noemaobs}, we describe the observations and the data reduction. In Sect.~\ref{sec:results}, we report and discuss our results on the \cii\ and dust continuum properties, SFR, and mass content of this luminous host galaxy. In Sect.~\ref{sec:conclusion}, we summarise our findings.

The magnitudes reported in this work are all in the AB system. Throughout the paper, we use a flat $\Lambda$ cold dark matter ($\Lambda$CDM) cosmology with $H_0$ = 70\,km\,s$^{-1}$\,Mpc$^{-1}$, $\Omega_\mathrm{m}$ = 0.3, and $\Omega_{\Lambda}$ = 0.7. All uncertainties are reported at the level of 1$\sigma$.

\section{\label{sec:noemaobs}NOEMA observations and data reduction }
We observed EUCL\,J1253$+$7054 in two visits, on 26 and 31 March, 2025, as part of the program E24AH (PIs: Decarli and Belladitta). The NOEMA array was in 12B configuration (resolution $\sim$0\farcs45) on 26 March and in 12C configuration (resolution $\sim$0\farcs85) on 31 March. The quasar 1044$+$719 served as phase and amplitude calibrator and MWC349 was observed as absolute flux calibrator. The quasars 3C273, 1448+762 and 0851+202 served as ancillary calibrators for bandpass and flux. The target was always observed at altitude $>45^\circ$. The system temperature was 150--250\,K on 26 March, and 100--140\,K on 31 March. The precipitable water vapour column was about 3\,mm during the first visit, and 1.2\,mm during the second visit.

\begin{figure*}[!ht]
    \centering
    {\includegraphics[width=0.45\linewidth]{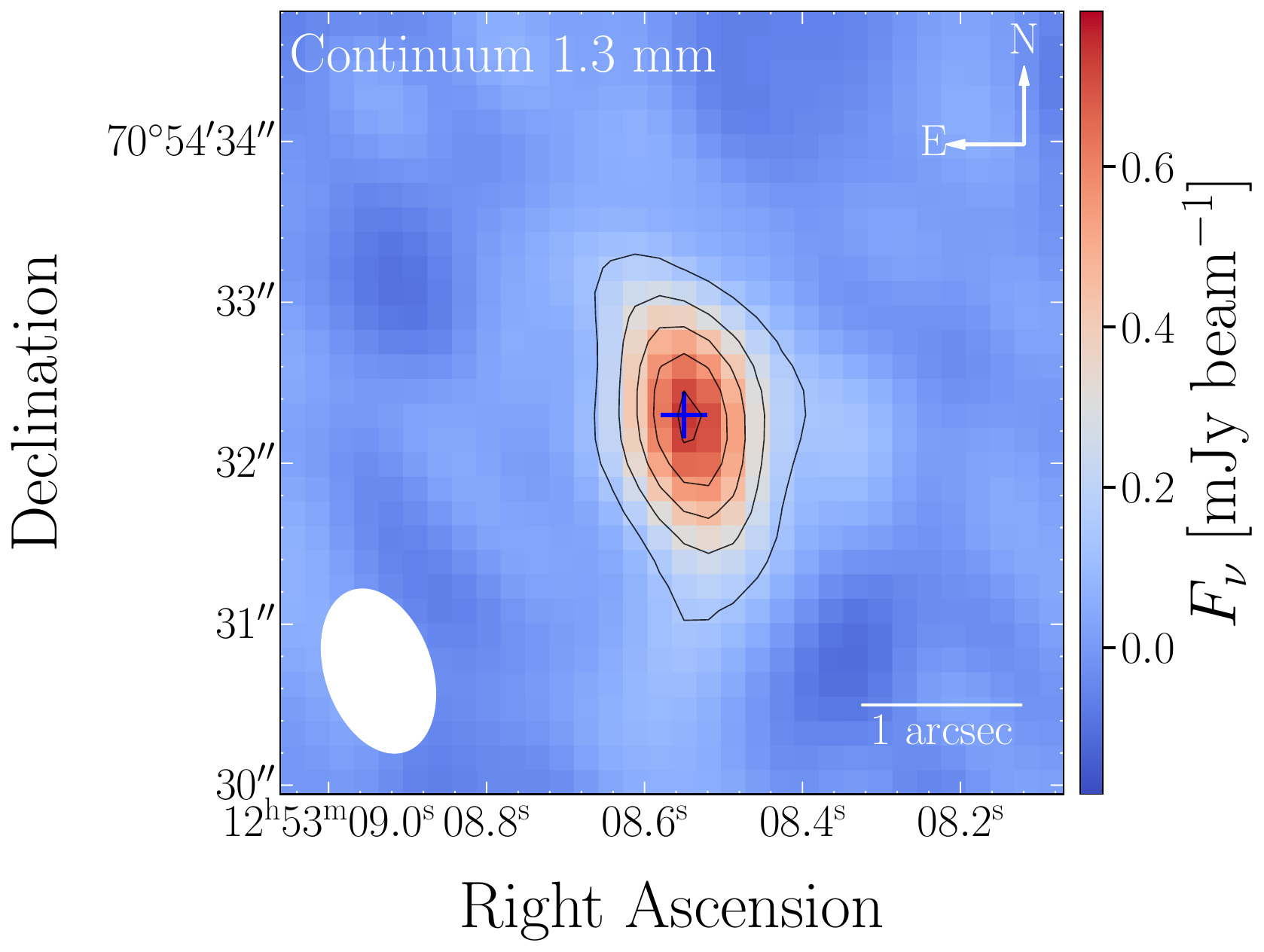}\hspace{0.1cm}
    \includegraphics[width=0.45\linewidth]{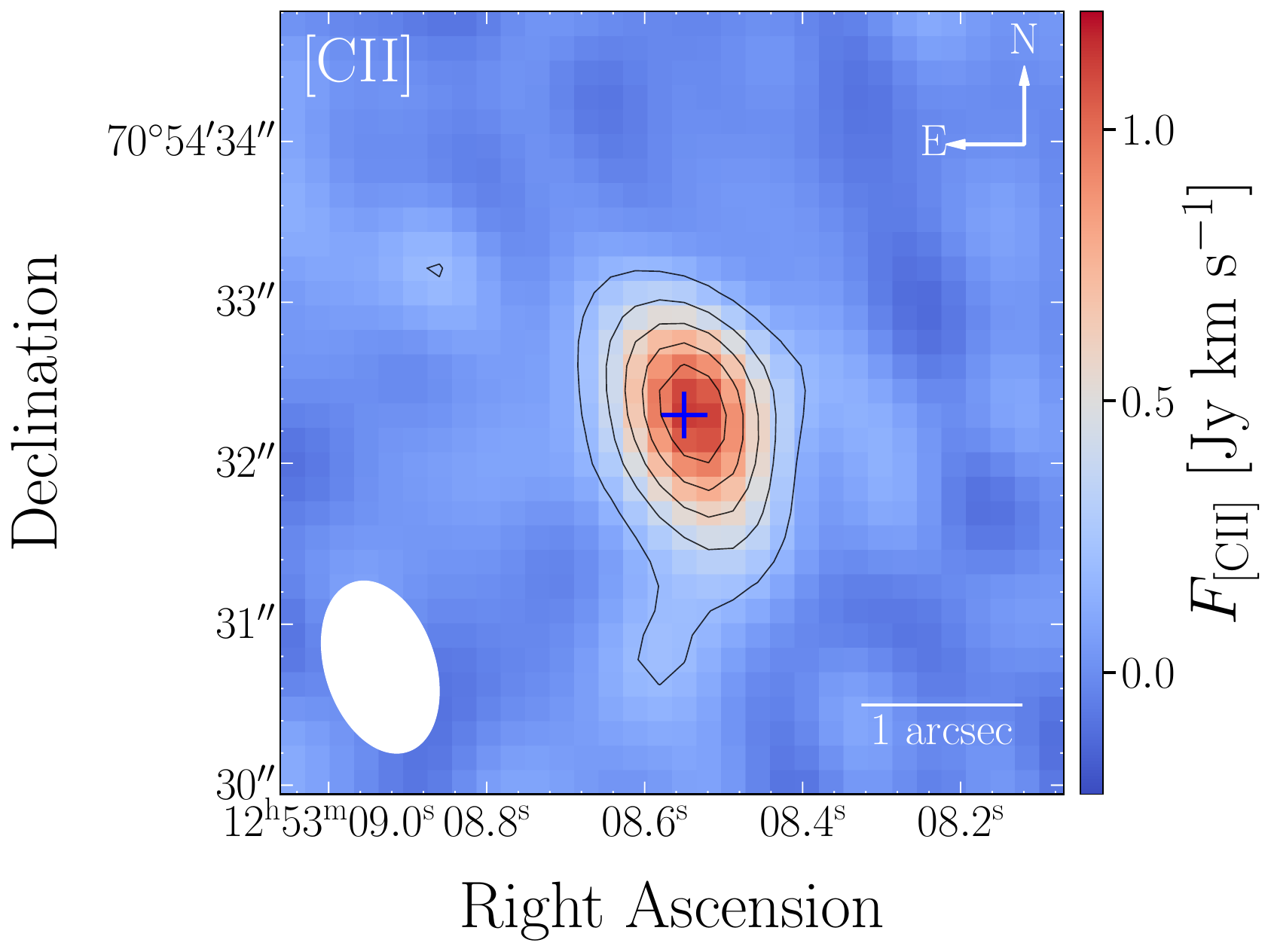}}
\caption{\ang{;0.08;}\,$\times$\,\ang{;0.08;} NOEMA 1.3\,mm continuum map (\textit{left}) and continuum-subtracted \cii\ map (\textit{right}) of the quasar EUCL\,J1253$+$7054. The blue cross shows the NIR position of the source from the \Euclid \JE-band image, consistent with the submm position. 
Contour levels correspond to 3, 6, 9, 12, and 15\,$\times$\,rms (0.05 and 0.2\,mJy\,beam$^{-1}$ for the continuum and the \cii\ map, respectively). The synthesised beam is plotted in the bottom left corners. North is up, east is left.}
    \label{fig:1mmcontinuum}
\end{figure*}

We calibrated the data using standard procedures, using the pipeline in the \textsf{clic} software offered within the Grenoble Image and Line Data Analysis Software
\citep[\textsf{GILDAS}\footnote{\url{https://www.iram.fr/IRAMFR/GILDAS/}};][]{gildas2013} suite. Only a small fraction of visibilities ($<25$\%, mostly from the longer baseline data obtained on 26 March) were flagged due to poor phase residuals. The calibrated dataset includes 22\,200 visibilities on source, corresponding to 4.20 hours (12 antennas equivalent). 
We processed the upper sideband (USB; 229.9--237.6\,GHz) and lower sideband (LSB; 214.3--222.1\,GHz) independently. 
For the \cii\ line analysis, we created a spectral cube from the LSB data, resampling the spectral axis in bins of 30\,km\,s$^{-1}$. 
As the USB is entirely free of line contamination, we produced a continuum image using all available USB channels. 
We note that the USB and LSB datasets (and their associated $uv$-tables) were kept separate throughout the analysis, and no combined USB$+$LSB continuum map was produced. Imaging was performed using \textsf{uvmap} in the \texttt{mapping} software. The resulting USB continuum map (left panel of Fig.~\ref{fig:1mmcontinuum}) has a beam size of 1\farcs05\,$\times$\,0\farcs66, PA = 18\,deg, and rms of 0.05\,mJy\,beam$^{-1}$. 

We created a continuum-subtracted \cii\ emission line cube using the \textsf{uv$\_$subtract} task in \texttt{mapping}. 
The line map was obtained by using the \textsf{uv$\_$average} task in \texttt{mapping}, and averaging over the range 218.364–218.814\,GHz for a total width of 420\,km\,s$^{-1}$, which corresponds to 1.9$\times$FWHM (see Table \ref{tab:results}). We show the \cii\ continuum-subtracted moment-0 map (rms = 0.20\,mJy\,beam$^{-1}$, beam size = 1\farcs1\,$\times$\,0\farcs7, PA = 17\,deg) in the right panel of Fig.~\ref{fig:1mmcontinuum}.

The following sections present our primary analysis conducted 
in the image plane. In addition, a complementary visibility-based analysis 
is provided in Sect.~\ref{app:uvfits} as a consistency check.

\begin{figure}
    \centering
    \includegraphics[width=1.0\linewidth]{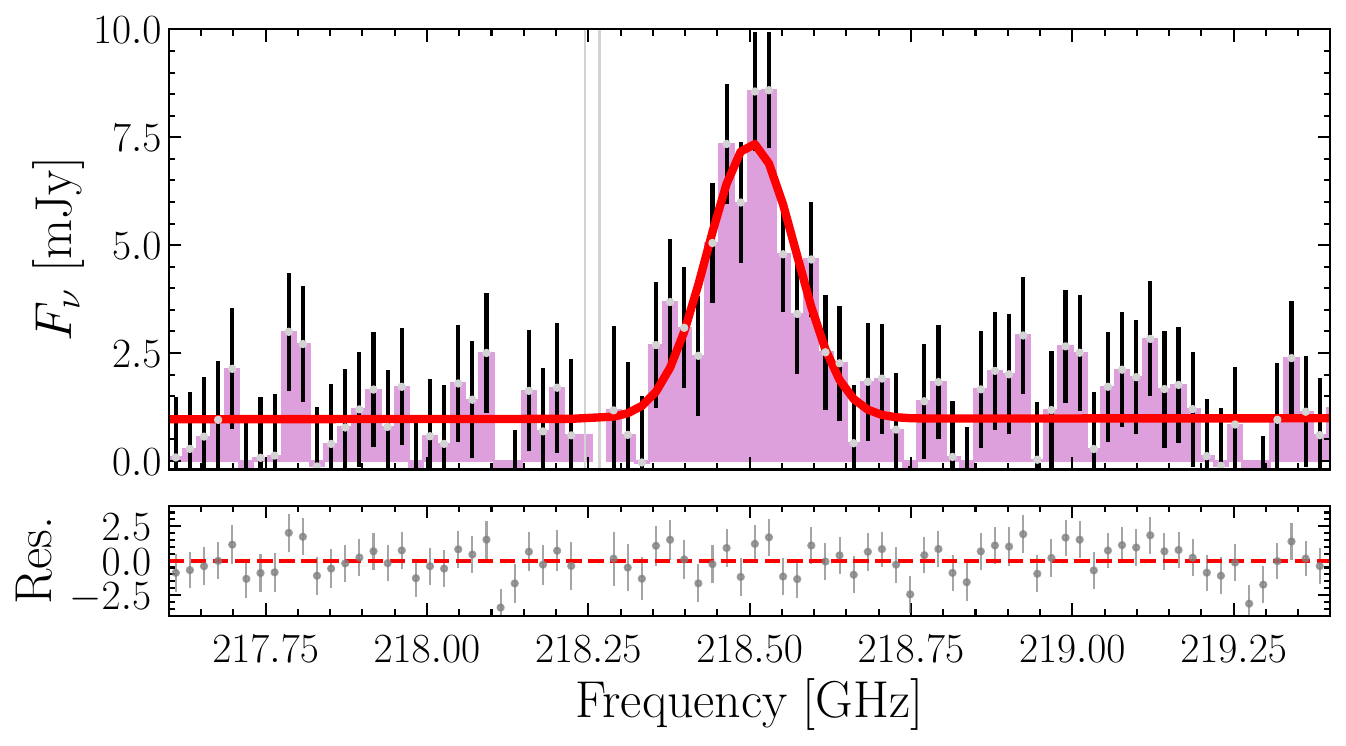}
    \caption{NOEMA spectrum of the \cii\ emission line in EUCL\,J1253$+$7054 (observed frame) obtained from the 30\,km\,s$^{-1}$ data cube. The red solid curve is the fit to the data consisting of a single Gaussian component plus continuum. A few channels close to the border of the sub-bands, near 218.25\,GHz have been masked (vertical grey lines). The bottom panel displays the residuals (data\,$-$\,model), with the dashed line marking the zero level.}
    \label{fig:cii_spec}
\end{figure}

\begin{figure*}
    \centering
    \includegraphics[]{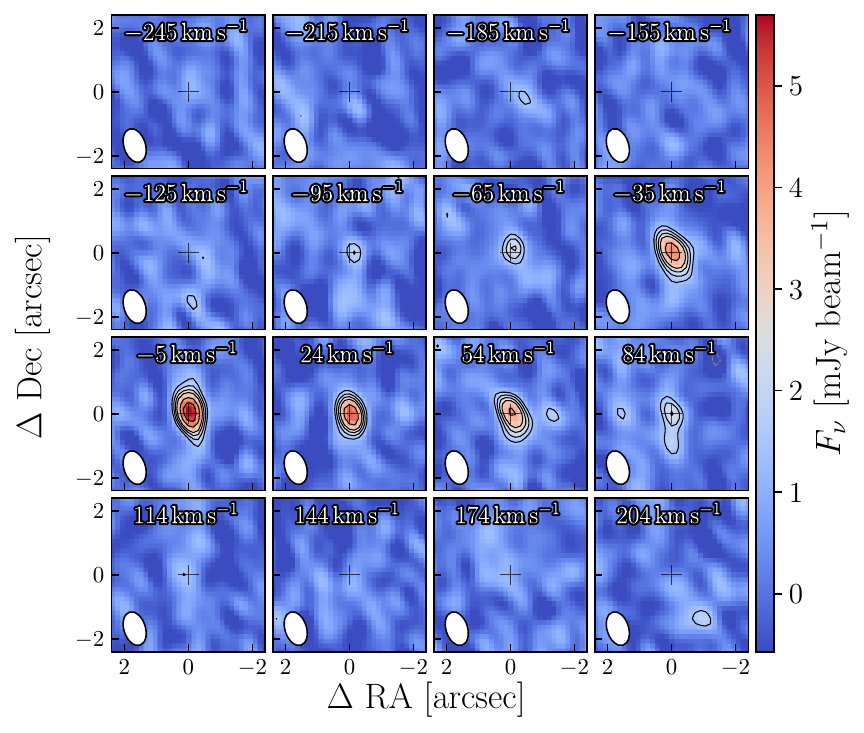}
    \caption{30\,km\,s$^{-1}$ \cii\ channel maps of the continuum-subtracted data of EUCL\,J1253$+$7054. Each channel is labelled with its central velocity. 
    The central black cross indicates the NIR position (same as Fig.~\ref{fig:1mmcontinuum}). Contours are plotted starting from 3$\sigma$, with steps of 1$\sigma$, where $\sigma$ is the noise in each individual channel. The synthesised beam is shown in the lower left corner of each panel. North is up, east is left.}
    \label{fig:channel}
\end{figure*}

\section{\label{sec:results}Results and discussion}

\subsection{\label{sec:ciresults} {\rm \cii} line measurements}
From the LSB data cube, we extracted the one-dimensional (1D) spectrum for the source in an aperture centred on the quasar position \citep[][]{novak2019,rojasruiz2021}. 
We determined the best aperture following \cite{mazzucchelli2025} by extracting the flux of the quasar in concentric apertures from 0\farcs1 to 3\farcs0\ (in steps of 0\farcs1) and choosing the radius at which the flux encounters a plateau. The best aperture found was $r = 1\farcs5$. We then fit the spectrum with a flat dust continuum emission and a Gaussian line. The extracted spectrum is shown in Fig.~\ref{fig:cii_spec} and we report the best fit values in Table~\ref{tab:results}.

As shown in Fig.~\ref{fig:cii_spec}, a single Gaussian component provides an excellent fit to the \cii\ line profile, yielding a reduced $\chi^2_{\nu} \approx 0.89$. 
We inspected the residuals for any systematic deviations that might indicate the presence of broad wings, which could trace outflowing gas. The residual spectrum is consistent with Gaussian noise across the entire line profile, with no evidence for additional components.
To further quantify this result, we performed a formal model comparison using the Bayesian information criterion (BIC) to test for the presence of a second Gaussian component. 
The double-Gaussian model was shown to be disfavoured by the data, yielding a $\Delta\mathrm{BIC} = 14.9$ relative to the single-component fit. 
Consequently, we adopted a single Gaussian profile as the sufficient and preferred model for the \cii\ emission.

The Gaussian fit on the 1D spectrum yields a \cii-based redshift of $7.6980\pm0.0004$.
We estimated the \cii\ line luminosity from the measured integrated line flux density following \cite{carilli2013},

\begin{equation}
  \frac{L_{\cii}}{\lsun} = 1.04 \times 10^{-3} \frac{F_\mathrm{line}}{\rm Jy\,km\,s^{-1}} \frac{\nu_\mathrm{obs}}{\rm GHz} \left(\frac{D_\mathrm{L}}{\rm Mpc}\right)^2,
  \label{eq:lcii}
\end{equation}

\noindent where $\nu_\mathrm{obs}$ is the observed frequency of the line and $D_\mathrm{L}$ is the luminosity distance. 

The line flux derived from the 1D Gaussian fit of the spectrum in Fig.~\ref{fig:cii_spec} is 1.51$^{+0.19}_{-0.17}$ Jy\,km\,s$^{-1}$ (Table~\ref{tab:results}).
This value is consistent with that derived from a 2D Gaussian fit performed with the task \texttt{IMFIT} of the Common Astronomy Software Applications package \citep[CASA,][]{mcmullin2007}, which yields an integrated flux density of ($3.80\pm0.35$)\,mJy, corresponding to ($1.59\pm0.15$)\,Jy\,km\,s$^{-1}$ when multiplying by the 420\,km\,s$^{-1}$ channel width used to create the line map (Sect.~\ref{sec:noemaobs}).
It is also consistent within 1$\sigma$ with the \cii\ flux density obtained from uv-plane fit (Sect.~\ref{app:uvfits}).
From the \cii\ flux and Eq.~(\ref{eq:lcii}) we derive a luminosity ($L_{\cii}$) of 2.03$^{+0.26}_{-0.23} \times 10^9\,\lsun$, making EUCL\,J1253$+$7054 the brightest \cii\ emitter among the known quasars at $z\approx7.5$ \citep{venemans2017,yang2020,wang2021}.
The IMFIT task indicates that the emission is marginally resolved, though with large uncertainties. 
The derived deconvolved \cii\ 
radius is $R_{\cii}$ = ($1.69\pm0.45$)\,kpc.
This value is consistent with those obtained by performing a Gaussian fitting directly in the uv-plane: ($1.07\pm0.15$)\,kpc (see Sect.~\ref{app:uvfits}.)
However, since the uv-plane analysis provides a more precise constraint (i.e. smaller uncertainty), we adopted this value as the reference for the dynamical mass calculation (see Sect.~\ref{sec:massbudget}).

We also computed the rest-frame equivalent width (EW), defined as the ratio between the flux of the line and the flux density of the continuum, finding a value of 1330$^{+200}_{-180}$\,km\,s$^{-1}$ (0.70$^{+0.10}_{-0.10}$\,\micron). This is consistent with the majority of the values of other $z>7$ quasars \citep[e.g.][]{wang2024}. 

Figure \ref{fig:channel} presents the continuum-subtracted \cii\ channel maps at 30\,km\,s$^{-1}$ resolution, which reveal the velocity structure of the source. 
The emission is concentrated within the central velocity channels, suggesting that the quasar host is not part of a merging or interacting system, unlike other $z>6$ quasars \citep[e.g.][]{decarli2017,banados2019,neeleman2021,izumi2021a,izumi2024}. 
We note a tentative southern extension at $\sim3\sigma$ significance in the 84\,km\,s$^{-1}$ channel map, although this feature is present in only a single velocity channel. 
While this could indicate a substructure at this velocity, consistent with the source being marginally resolved, 
deeper and higher angular-resolution observations are needed to better constrain its kinematic properties.

\subsection{\label{sec:continuum} Continuum measurements}
We derive the flux density of the continuum at 1.3\,mm (observed frame) from both the 1D spectral fit and the 2D fit of the continuum map.
The 2D Gaussian fit on the USB continuum map was performed with CASA/\texttt{IMFIT}.
We obtain an integrated flux density of ($1.17\pm0.09$)\,mJy, consistent with the value computed from the 1D fit on the LSB data cube (see Table~\ref{tab:results}) and a peak flux density of ($0.70\pm0.05$)\,mJy\,beam$^{-1}$. 
The derived continuum flux is consistent within 1$\sigma$ with the value obtained from fitting directly in the uv-plane (Sect.~\ref{app:uvfits}).

In addition to the extraction of the flux density, we also find the following results: (i) the coordinates of the 1.3\,mm continuum emission (RA = \ra{12;53;08.5409}, Dec = \ang{+70;54;32.2545}) are only 0\farcs06 from the coordinates based on the \Euclid NISP \citep{EuclidSkyNISP} \JE-band image \citep{yang2026}, which is within the astrometric uncertainty of NOEMA; and (ii) the uv-plane analysis (Sect.~\ref{app:uvfits}) indicates that the source is resolved, with a measured full width at half power of $0\farcs45\pm0\farcs04$ (assuming a circular Gaussian profile). This is consistent with the image-plane estimates, where the ratio between integrated and peak flux densities ($\sim$\,1.7) suggests that the emission is spatially extended.

To infer the IR luminosity, we modelled the dust continuum emission as a modified black body (MBB) with a dust temperature of
$T_\mathrm{dust}=47$\,K and an emissivity index of $\beta = 1.6$ \citep[e.g.][]{beelen2006}. 
We adopted these fiducial values following the standard approach used in submm studies of high redshift quasars \citep[e.g.][]{banados2015,decarli2018,novak2019,venemans2018,khusanova2022,mazzucchelli2025}, ensuring our results are directly comparable with the literature.
This choice is also consistent with the median $T_\mathrm{dust}= (48\pm3)$\,K derived by \cite{costa2026} from detailed SED fitting to a sample of 11 high-redshift quasars with well-sampled photometry.
Similarly, our adopted $\beta=1.6$ is consistent with empirical measurements of $\beta= (1.8\pm0.3)$ reported by \cite{witstok2023} for high-redshift objects.

Given the high redshift nature of EUCL\,J1253$+$7054, we also take into account the effect of the cosmic microwave background (CMB) using the correction prescription of \cite{dacunha2013}.
We also assumed the opacity law of \cite{dunne2000}: $k_{\nu} = 0.77[\nu/(352\,\mathrm{GHz})]\mkern 1mu ^{\beta}$\,cm$^2$\,g$^{-1}$.  
We then computed the IR luminosities by integrating the MBB function between different frequencies: $L_\mathrm{FIR}$ from 60 to 100\,\micron\ \citep{helou1988} and $L_\mathrm{TIR}$ from 8 to 1000\,\micron\  \citep[e.g.][]{sanders2003}.
The results are reported in Table~\ref{tab:results}. 
We note that these results, as well as the derived SFR (see Sect. \ref{sec:sfr}) and \cii\ deficit (see Sect. \ref{ciideficit}) values, are sensitive to the adopted values of $T_\mathrm{dust}$ and $\beta$, which are known to differ in high-$z$ sources \citep[e.g.][]{leipski2013,venemans2018,tripodi2023,izumi2024}.
Assuming the range of dust parameters explored by \cite{costa2026} in their analysis of a sample of 11 high-redshift quasars ($T_\mathrm{dust}=34$--65\,K and $\beta=2.0$--2.2), we find that the values we report vary by factors of $\approx2-5$; however, they still remain consistent with those typically observed in the high-redshift quasar population. 
We further tested the impact of optical depth, $\tau$, effects by adopting an optically thick modified blackbody model. 
Well-sampled submm SED fits have revealed that $z>6$ quasars can have non-negligible dust optical depths, with values reaching $\approx0.5$ \citep[e.g.][]{decarli2023,costa2026}. 
As an illustrative case, we assumed $\tau=1$ at the rest frame frequency of the \cii\ line and found that the inferred IR luminosities decrease by a factor of $\approx2.6$, while the dust mass (reported in Sect. \ref{sec:massbudget}) increased by $\approx0.2$ dex. 
\cite{venemans2018} found similar variations when performing comparable tests.
These variations propagate to derived quantities such as SFR$_{\rm IR}$ and the \cii\ deficit, but the resulting values remain well within the range observed for high-$z$ quasar host galaxies. 
This indicates that the uncertainties on the derived parameters are dominated by systematic effects, rather than by statistical errors.

\begin{table}[!ht]
    \centering
        \caption{Observed and derived properties of EUCL\,J1253$+$7054.}
        \label{tab:results}
    \begin{tabular}{lr}
    \hline\hline\noalign{\vskip 2pt}
    $\nu_\mathrm{obs,\,\cii}$ [GHz] &  $218.503\pm0.009$ \\[1pt]
    $z_{\cii}$ &  $7.6980\pm0.0004$ \\[1pt]
    $F_{\cii}$ [Jy\,km\,s$^{-1}$] &  1.51$^{+0.19}_{-0.17}$ \\[1pt]
    $F_{\nu}$\,(cont.) [mJy] &  $1.13\pm0.08$ \\[1pt]   
    FWHM$_{\cii}$ [km\,s$^{-1}$]   & 222$_{-32}^{+44}$ \\[2pt]
    EW$_{\cii}$ [km\,s$^{-1}$] & 1330$^{+200}_{-180}$ \\ 
    
\noalign{\vskip 2pt}
\hline\noalign{\vskip 2pt}
    $L_{\cii}$ [$\lsun$] & 2.03$^{+0.26}_{-0.23}\,\times\,10^9$ \\[1pt]
    $L_{\rm FIR}$ [$\lsun$] & $3.6\pm0.3\,\times\,10^{12}$ \\[1pt] 
    $L_{\rm TIR}$ [$\lsun$] & $4.8\pm0.3\,\times\,10^{12}$ \\[1pt] 
    SFR$_{\rm IR}$ [$\msun$\,yr$^{-1}$] & $715\pm53$ \\[1pt] 
    SFR$_{\cii,\,\rm DL14}$ [$\msun$\,yr$^{-1}$] & 289$^{+45}_{-39}$ \\[1pt]  
    SFR$_{\cii,\,\rm HC15}$ [$\msun$\,yr$^{-1}$] &  644$^{+86}_{-76}$ \\[1pt]  
    $M_\mathrm{dust}$ [$\msun$] &  $1.4\pm0.1 \times 10^{8}$ \\[1pt] 
    $M_\mathrm{gas}$ [$\msun$] & 0.41$-$1.24 $\times$ $10^{10}$ \\    
    $M_\mathrm{dyn}$ [$\msun$] (Eq.~\ref{eq:mdyn1}) &  3.3$^{+1.4}_{-1.1}\times 10^{9}$ \\[1pt]
    $M_\mathrm{dyn}$ [$\msun$]  (Eq.~\ref{eq:mdyn2}) &  1.3$^{+0.6}_{-0.4}\times10^{10}$ \\   

\noalign{\vskip 2pt}
    \hline
    \end{tabular}
    \begin{tablenotes}
    \small
    \item \textbf{Notes:} Col (1) \cii~line peak frequency, \cii~ redshift, integrated line flux and continuum flux density at 1.3\,mm, \cii\ FWHM, \cii\ EW value; \cii\ luminosity, IR luminosities, SFR inferred from \cii\ line and from the IR luminosity, mass of the dust and molecular gas and dynamical mass. Col (2) Parameter values. All parameters assume an optically thin dust model with $T_\mathrm{dust} = 47$\,K and $\beta = 1.6$ for direct comparisons with the literature. See text in Sect. \ref{sec:continuum} for alternative assumptions.
    \end{tablenotes}
\end{table}

\subsection{UV-plane analysis of the continuum and \cii\ line emission}
\label{app:uvfits}
For completeness, we report here the results of Gaussian fitting performed directly in the uv-plane using the \textsf{uv\_fit} task in \textsf{GILDAS}, adopting a circular Gaussian profile. 
We present the flux densities of the 1.3\,mm continuum 
and the \cii\ line emission, as well as the source size derived from the \cii\ emission.

We obtained a \cii\ integrated flux density of $F_{\rm [CII]} = (3.56 \pm 0.25$)\,mJy, which corresponds to ($1.49\pm0.10$)\,Jy\,km\,s$^{-1}$,
and a continuum flux density of $F_{\nu}\,\mathrm{(cont.)} = (1.06 \pm 0.06$)\,mJy. 
From the uv-plane fit, we derived a \cii\ FWHM of $0\farcs43 \pm 0\farcs06$, 
corresponding to a physical diameter of ($2.14 \pm 0.30$)\,kpc (radius $1.07 \pm 0.15$\,kpc) at $z = 7.698$. 
All measurements are consistent within $1\sigma$ with the values derived from the image-plane analysis (Sects.~\ref{sec:ciresults} and \ref{sec:continuum}), confirming our results.

\subsection{\label{ciideficit} {\rm \cii} deficit}
The \cii/IR luminosity ratio serves as a diagnostic of ISM properties, and this ratio typically ranges from 0.003 to 0.01 in local star-forming galaxies with modest dust temperatures and luminosities \citep[see][for a recent review]{hodge2020}. 
However, this ratio systematically decreases by up to an order of magnitude in more IR-luminous systems ($L_\mathrm{IR} > 10^{11}\,\lsun$), such as LIRGs and ULIRGs, a phenomenon known as the \cii\ deficit \citep[e.g.][]{farrah2013,diazsantos2017}. 
Similar deficits have been observed in other FIR lines, including \oi\ 63.2\,\micron, \oi\ 145\,\micron, and \nii\ 122\,\micron, leading to the broader term `FIR line deficit'  \citep{gracia2011}. 
The physical origin of this trend is still debated \citep[see][for a recent review]{decarli2025}, and is not simply a function of IR luminosity or redshift \citep[e.g.][]{magdis2014}.
Several proposed explanations include AGN-driven ionisation of C$^+$ \citep[e.g.][]{langer2015}, self-absorption in dense environments \citep[e.g.][]{diazsantos2017}, thermal saturation under intense far-UV radiation \citep[e.g.][]{rybak2019}, and optically thick \cii\ emission associated with compact, high-temperature dust \citep[e.g.][]{casey2014}. 

\begin{figure}[!ht]
    \centering   \includegraphics[]{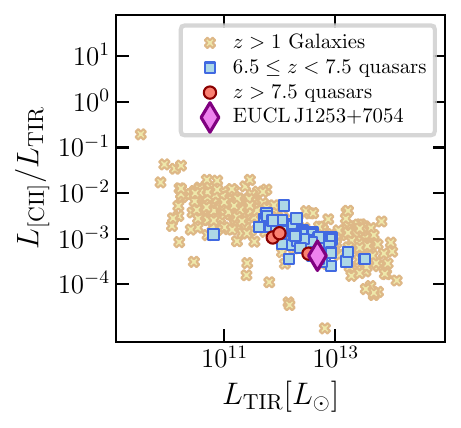}
    \caption{\cii-to-IR luminosity ratio as a function of IR luminosity for EUCL\,J1253$+$7054, compared with measurements from the literature: high-redshift quasars are from \cite{wang2024} and \cite{bouwens2025}, while the compilation of galaxies is from \cite{decarli2025} and include ULRIGs, Submillimeter galaxies, optically selected galaxies (primarily Lyman-break galaxies and
Ly$-\alpha$ emitters) starbursts, main-sequence galaxies \citep[see][for all the details]{decarli2025}.} 
    \label{fig:ciideficit}
\end{figure}

The \cii\ deficit of EUCL\,J1253$+$7054 ($L_\mathrm{\cii}/L_\mathrm{TIR}=4.2\times10^{-4}$) is shown in Fig.~\ref{fig:ciideficit}. 
This is in line with the behaviour of the high-$z$ quasar population and local IR-luminous galaxies, which display $L_\mathrm{\cii}/L_\mathrm{TIR}$ ratios ranging from ULIRG-like values to those of normal star-forming galaxies. 
This indicates that the \cii\ deficit is not strongly affected by the presence of a luminous active black hole \citep[e.g.][]{magdis2014,diazsantos2013,decarli2018,decarli2025,venemans2020,wang2024}.
We note that varying $T_\mathrm{dust}$ and $\beta$ (see Sect.~\ref{sec:continuum}) changes the inferred \cii\ deficit by factors of $\approx 5$ (lower) to $\approx 1.5$ (higher) relative to our fiducial assumptions. 
However, all cases remain consistent with the \cii\ deficit observed for high-redshift quasars and luminous IR sources.

The observed \cii\ deficit in EUCL\,J1253$+$7054 provides insights beyond strictly what the morphological analysis alone can reveal. \cite{herrera2018a} demonstrated that at a fixed IR luminosity, the \cii/IR ratio decreases as galaxies become more compact. 
Our measured \cii/IR ratio of 4.2\,$\times\,10^{-4}$, combined with the high FIR luminosity, suggests that EUCL\,J1253$+$7054 is compact, with high-surface-density star formation.

The $x$-axis of Fig.~\ref{fig:ciideficit} shows that EUCL\,J1253$+$7054 exhibits the highest IR continuum luminosity among the three $z>7.5$ known quasars, though comparable to J1007$+$2115 \citep[][]{wang2024}. 
However, unlike the \cii\ luminosity, which is a robust measurement, IR luminosities are more model-dependent due to assumptions about dust temperature and emissivity (see Sect. \ref{sec:continuum}).

\subsection{\label{sec:sfr} Star-formation rate}
We can use both the luminosity from the \cii\ line and the dust continuum to infer the SFR of the host galaxy of EUCL\,J1253$+$7054. 
We followed the methodology reported in e.g.\ \citet{decarli2018} and \citet{mazzucchelli2023}.
Compared to the IR luminosity, \cii\ offers key observational advantages: it is bright and readily detectable, ubiquitous in galaxies, facilitates accurate redshift measurements, and is rarely saturated except in dense starburst nuclei.
However, the \cii\ deficit in luminous sources and its metallicity dependence limit its reliability as an SFR tracer for very bright, compact starbursts and metal-poor systems \citep[e.g.][hereafter DL14 and HC15]{delooze2014,herrera2015}. 
Conversely, dust continuum photometry enables SFR estimates without precise redshift information \citep{kennicutt2012}, but requires sampling near the SED peak ($\sim$100\,\micron\ rest-frame), which is observationally challenging from the ground due to poor atmospheric transmission below 500\,GHz.

In this case, we can use the equation from \cite{kennicutt2012} to estimate the dust-based SFR (SFR$_\mathrm{IR}$) via
\begin{equation}
        \frac{\rm SFR_{IR}}{\rm \msun\,yr^{-1}} = 1.49\times10^{-10} \left(\frac{L_\mathrm{IR}}{\lsun}\right), \label{eq:sfrdust}
\end{equation}
where the IR luminosity ($L_\mathrm{IR}$) is computed between 3 and 1100\,\micron. 
The \cite{kennicutt2012} relation assumes that the FIR emission is primarily powered by star formation. 
This approach is consistent with observational studies of high-$z$ quasars \citep[e.g.][]{barnett2015,decarli2023,costa2026} which find that cold dust emission in $z > 6$ quasar hosts is aptly modelled by intense starburst activity. 
However, we note that the presence of a powerful AGN may provide an additional heating source for the dust. 
While several studies have found this contribution to be minor in the FIR \citep[e.g.][]{leipski2014,venemans2017_ir,neeleman2019,li2022}, some theoretical models suggest the central SMBH could contribute a non-negligible fraction of the IR luminosity \citep[e.g.][]{schneider2015}. 
In such a scenario, our IR-based SFR would be overestimated.

We computed the \cii-based SFR (SFR$_{\cii}$) from the relation reported in Eq.~(17) of DL14, calibrated on $z=0.5$--6.6 galaxies with a scatter of about 0.4\,dex, via
\begin{equation}
     \frac{\rm SFR_{\cii}}{\rm \msun\,yr^{-1}} = 3\times10^{-9} \left( \frac{L_{\cii}}{\lsun} \right)^{1.18}. 
     \label{eq:sfrcii_delooze}
\end{equation}

We obtained SFR$_\mathrm{IR}$ = ($715\pm53$)\,$\rm \msun\,yr^{-1}$ and SFR$_{\cii,\,\rm DL14}$ = 289$^{+45}_{-39}\,\rm \msun\,yr^{-1}$.
These estimates are also listed in Table~\ref{tab:results}. 
These values are in line with the SFRs of $z>6$ quasars reported in the literature using the same estimators \citep[e.g.][]{venemans2017,yang2020,wang2024}. 

The \cii-based SFR is commonly observed to be lower than the FIR-based SFR in FIR-luminous quasar host galaxies at $z>6$ \citep[e.g.][]{wang2024}. 
This discrepancy arises because the empirical relation from DL14 was calibrated on typical star-forming galaxies, whereas FIR-luminous galaxies  exhibit a significant \cii\ deficit at high luminosities (see Sect. \ref{ciideficit}). 
HC15 provided a relation for deriving the SFR from the \cii\ luminosity that accounts for the \cii\ deficit via
\begin{equation}
\frac{\rm SFR_{\cii}}{\rm \msun\,yr^{-1}} = 0.052 \left( \frac{L_{\cii}\,\Psi(y)}{10^{40}\,\text{erg\,s}^{-1}} \right)^{1.034}, 
\label{eq:sfrcii_hc}
\end{equation}
where $\Psi(y) = (y/y_\mathrm{t})^{\alpha}$ is a colour adjustment coefficient dependent on dust temperature, with $y = \nu F_\nu(70\,\micron)/\nu F_\nu(160\,\micron)$, $y_\mathrm{t} = 1.12$ and $\alpha = 1.2$ (see HC15 for details).
Applying this equation, we derived SFR$_{\cii,\,\rm HC15}$ = 644$^{+86}_{-76}\,\rm \msun\,yr^{-1}$, which is in good agreement with the FIR-based estimate. 

\subsection{Mass budget: $M_\mathrm{dust}$, $M_\mathrm{gas}$, $M_\mathrm{dyn}$, and $M_\mathrm{BH}$}
\label{sec:massbudget}

The dust mass, $M_\mathrm{dust}$, is typically derived by scaling the observed continuum flux density to a MBB template normalised to 1\,$\msun$ of dust, following the standard optically thin assumption whereby the dust emission is directly proportional to dust mass \citep[e.g.][]{dacunha2013}. 
We employed the same method and assumptions described in Sect. \ref{sec:continuum} (i.e. $T_\mathrm{dust}=47$\,K and $\beta = 1.6$), finding a value of $\logten(M_\mathrm{dust}/\msun)=8.14\pm0.03$, which is consistent with the dust mass values derived for several quasars at $z>6.5$ \citep[e.g.][]{wang2024}. 

Dust emission can also be used as a molecular gas tracer. By assuming a gas-to-dust ratio, $\delta_{\rm GDR}$, of 100, which is a value that yields consistent results across different gas-mass tracers \citep[e.g.][]{decarli2022,kaasinen2024}, and a molecular-to-total gas mass fraction of 0.75 \citep[e.g.][]{neeleman2021,decarli2023}, we can roughly estimate the molecular gas mass ($M_\mathrm{gas}$) to be around 10$^{10}\,\msun$. 

We caution that the gas mass estimate is sensitive to the assumed 
$\delta_{\rm GDR}$, which is an uncertain quantity \citep[e.g.][]{feruglio2023,tripodi2024}. 
Adopting the range $\delta_{\rm GDR} = (80 \pm 40)$ \citep{salvestrini2025}, the molecular gas mass varies between $\approx4.1\times10^9 \msun$ and $\approx1.2
\times10^{10} \msun$. 

Combining the molecular gas mass with the SFR derived from the IR luminosity, we obtained a gas depletion time of $\approx8-20$\,Myr, in line with values reported for other $z>6$ quasar hosts \citep[e.g.][]{kaasinen2024,tripodi2024}. 
This suggest that the host galaxy of EUCL\,J1253$+$7054 is undergoing an extreme starburst phase, as commonly observed in IR-luminous $z>6$ quasar hosts.

From these NOEMA observations, we can also derive a first estimate of the dynamical mass ($M_\mathrm{dyn}$) of the system.  
In our case, the FIR continuum and \cii\ emission of EUCL\,J1253$+$7054 host galaxy are marginally unresolved (see Fig.~\ref{fig:1mmcontinuum}), preventing us from performing a full modelling of the kinematics of our source; namely, we are not able to distinguish between ordered rotation or more complex kinematics \citep[e.g.][]{neeleman2021}. 
Therefore, we estimate $M_\mathrm{dyn}$ in a simplified way, following \citet{decarli2018}.
Under the assumption of a dispersion-dominated system, we can express $M_\mathrm{dyn}$ with the equation
\begin{equation}
    M_\mathrm{dyn} = \frac{3}{2} \frac{R_{\cii} \sigma_{\cii}^2}{G},
    \label{eq:mdyn1}
\end{equation}    
where $R_{\cii}$ is the size of the \cii-emitting region, defined from a Gaussian fit of the visibilities ($1.07\pm0.15$\,kpc, see Sect.~\ref{app:uvfits}), $\sigma_{\cii}$ is derived from the Gaussian fit of the line profile (94.5$^{+18.6}_{-13.5}$\,km\,s$^{-1}$) and $G$ is the gravitational constant.

\begin{figure}[!ht]
    \centering
    \includegraphics[width=1\linewidth]{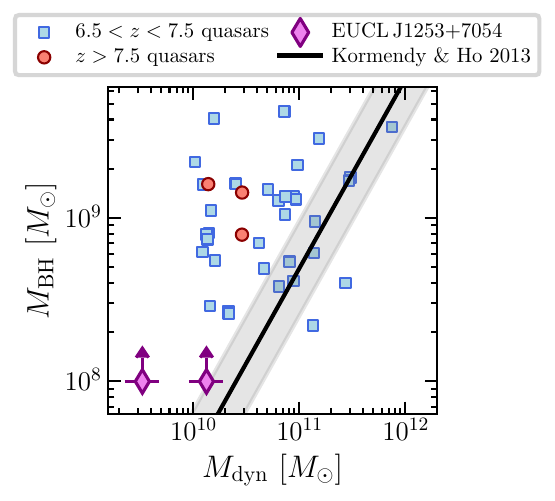}
    \vspace{-0.5cm}
    \caption{$M_\mathrm{BH}$ versus $M_\mathrm{dyn}$ of EUCL\,J1253$+$7054 compared to measurements from other high-$z$ quasars \citep[data from][]{wang2024}. For our source, we plot two points at the two estimated $M_\mathrm{dyn}$ values (see Sect.~\ref{sec:massbudget}) and at the lower limit of the black hole mass (assuming Eddington-limited accretion). The black line represents the local relation of \citet{kormendy2013}.}
    \label{fig:mbhmdyn}
\end{figure}

In the cases where the line width is dominated by rotation, the gas assumes a flat disk structure with inclination angle $i$ \citep[e.g.][]{willott2015}. 
Under this assumption, the dynamic mass can be expressed as 
\begin{equation}
M_\mathrm{dyn} = \frac{R_{\cii}}{G} 
\left( \frac{0.75\,\mathrm{FWHM}_{\cii}}{\sin(\textit{i})} \right)^{2}.
    \label{eq:mdyn2}
\end{equation}   
We adopted an inclination angle of $i = 46^{\circ}$, which represents the median inclination of the $z>6$ quasar hosts sample of \citet{wang2024}.
By using both Eq.~(\ref{eq:mdyn1}) and Eq.~(\ref{eq:mdyn2}), we found the $M_\mathrm{dyn}$ value to be $3.3^{+1.4}_{-1.1}\times 10^{9}$\,\msun\ and $1.3^{+0.6}_{-0.4}\times10^{10}$\,\msun, respectively.
Previous studies \citep[e.g.][]{decarli2018} have shown that Eq.~(\ref{eq:mdyn2}) typically yields values that are three to four times higher than those obtained with Eq.~(\ref{eq:mdyn1}). 
For a discussion of the uncertainties involved in deriving dynamical masses from unresolved data, we refer to \citet{deblok2014}. 
The lower $M_\mathrm{dyn}$ estimate is one order of magnitude lower than the dynamical mass of the other $z\sim7.5$ quasars \citep[e.g.][and see Fig.~\ref{fig:mbhmdyn}]{venemans2017,wang2024}. 

The limited angular resolution of our data prevents a precise determination of the \cii\ extent and our $R_{\cii}$ estimate should be interpreted with caution. 
We cannot exclude the presence of more extended \cii\ emission, which would increase the derived dynamical mass, placing it closer to the bulk population of high-z quasars ($M_\mathrm{dyn}\gtrsim10^{11}$, see Fig.~\ref{fig:mbhmdyn}). 
However, extended \cii\ halos (on scales of $\approx$10\,kpc), as observed in $z=5$--7 star-forming galaxies \citep[e.g.][]{fujimoto2019}, have not been detected in $z>6$ quasar hosts to date, even in high-resolution ALMA observations \citep[e.g.][]{neeleman2021}. Our estimates of $M_\mathrm{dust}$, $M_\mathrm{gas}$, and $M_\mathrm{dyn}$ are all reported in Table~\ref{tab:results}.

When combined with SMBH mass measurements, $M_\mathrm{dyn}$ provides insight into SMBH--galaxy co-evolution (assuming that the total mass traced by the dynamical mass is dominated by stars and gas).  
Several studies have revealed that SMBHs in the EoR appear to be overmassive relative to their local Universe counterparts \citep[e.g.][]{venemans2016,shao2017,farina2022,wang2024}, with high-redshift quasars exhibiting higher $M_\mathrm{BH}/M_\mathrm{dyn}$ ratios than local galaxies.
However, this apparent evolution is strongly influenced by observational selection biases, since high-redshift observations primarily target the most luminous quasars, which preferentially sample galaxies hosting exceptionally massive black holes \citep[e.g.][]{volonteri2016}. 
Indeed, recent works \citep[e.g.][]{izumi2021b,bouwens2025} have shown that UV-faint quasars at $z>6$ could be hosted by very massive galaxies while simultaneously harbouring lower mass black holes ($<10^9$\,\msun), making the $M_\mathrm{BH}/M_\mathrm{dyn}$ ratio consistent with the Local Universe relation and a factor of $\sim$15 lower than the ratios observed in UV-bright quasars. 
Using detailed modelling, \citet{silverman2025} and \citet{li2025} suggest that when observational biases and measurement uncertainties are taken into account, the $M_\mathrm{BH}/M_\mathrm{dyn}$ ratios of high-redshift quasars actually follow the same black hole--galaxy mass relation as their local counterparts. This is consistent with the de-biased local relation of e.g., \citealt[][]{shankar2016}.

The optical/NIR data that we have in hand for EUCL\,J1253$+$7054 \citep[see][]{yang2026} do not cover the \civ\ and \mgii\ broad emission lines that are commonly used to estimate $M_\mathrm{BH}$ through the standard single epoch method \citep[e.g.][]{farina2022,belladitta2025}. However, assuming that the quasar radiates below or at the Eddington limit, we can estimate a lower limit on the black hole mass,
\begin{equation}
\frac{M_\mathrm{BH}^\mathrm{min}}{\msun} = \frac{L_\mathrm{bol}}{1.26 \times 10^{38}~\rm erg~s^{-1}}\;,
   \label{eq:bhmassedd}
\end{equation}
where $L_\mathrm{bol}$ is the bolometric luminosity of EUCL\,J1253$+$7054, with $\logten (L_\mathrm{bol}/\lsun) = 46.10$ reported in \cite{yang2026}, which was inferred from $M_{1450}$ using the empirical conversion from \citet{runnoe2012} and under the assumption of negligible obscuration. 
From this equation we obtain a value of $M_\mathrm{BH}^\mathrm{min}$ of 10$^8\,\msun$, consistent with black hole masses of UV-faint and bright $z>6$ quasars \citep[e.g.][]{willott2010,onoue2019,yang2021}. 
The estimated $M_\mathrm{BH}/M_\mathrm{dyn}$ ratio is in the range 0.007--0.03, due to large uncertainties in both measurements. 
This range of values is similar to that of the UV-luminous high-$z$ quasars reported in the literature \citep[e.g.][see Fig.~\ref{fig:mbhmdyn}]{wang2024}. A dynamical mass $>10^{10}\,\msun$ and a black hole mass below a few 10$^8\,\msun$  would be needed to place EUCL\,J1253$+$7054 on the local scaling relation.

\section{\label{sec:conclusion}Concluding remarks}

\begin{figure}
    \centering   \includegraphics[]{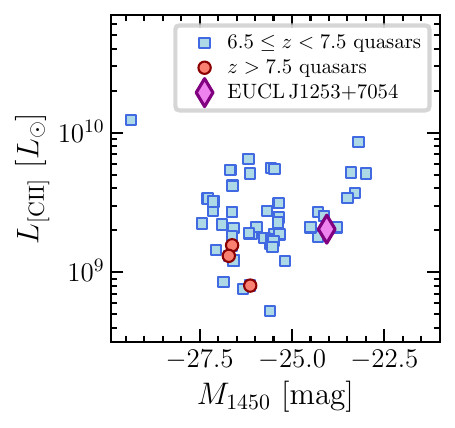}
    \caption{$L_{\cii}$ vs $M_{1450}$ for EUCL\,J1253$+$7054 and for $z\geq6.5$ quasars reported in the literature \citep{wang2024,bouwens2025,decarli2025}. Despite its lower UV brightness, EUCL\,J1253$+$7054 is the brightest \cii\ emitter at $z>7.5$, suggesting that the quasar UV luminosity is not a good proxy of the host-galaxy \cii\ emission.} 
    \label{fig:muvlcii}
\end{figure} 
In this paper we report NOEMA observations of the $z_{\cii}=7.6980\pm 0.0004$ quasar EUCL\,J1253$+$7054, revealing it to be the brightest \cii\ emitter among the $z>7.5$ quasars targeted to date (see Fig.~\ref{fig:muvlcii}), despite being nearly two magnitudes fainter in
the rest-frame UV ($M_{1450}=-24.06$).   
This finding aligns with the recent findings by \citet{bouwens2025}, who showed that UV-faint quasars at $z\sim6$  are predominantly hosted by massive, \cii-luminous star-forming galaxies. 
If quasars universally reside in massive galaxies, variations in their UV luminosity may primarily reflect dust obscuration \citep[e.g.][]{kato2020,izumi2021a} or extreme sub-Eddington accretion \citep[e.g.][]{matsuoka2019}, rather than differences in host galaxy mass. As a result, UV-faint quasars like EUCL\,J1253$+$7054 could represent a different evolutionary phase from UV-bright quasars, possibly marking a transition between overmassive black holes at early times and the local $M_\mathrm{BH}/M_\mathrm{dyn}$ relation.

Testing this scenario requires a larger sample of $z>7.5$ quasars spanning a range of UV luminosities, with both host galaxy measurements (e.g. with ALMA/NOEMA) and black hole mass constraints (e.g. with JWST). 
The \Euclid mission is poised to deliver such a sample in the coming years \citep{Barnett-EP5, banados2025euclid}, fundamentally changing the landscape of black hole studies in the early Universe.
 
%

\begin{acknowledgements}
We thank the anonymous referee for their valuable suggestions, which improved the quality and clarity of this paper.

\AckEC  

This work is based on observations carried out under project number E24AH with the IRAM NOEMA Interferometer. IRAM is supported by INSU/CNRS (France), MPG (Germany) and IGN (Spain).

R.~Decarli acknowledges support from the INAF GO 2022 grant ``The birth of the giants: JWST sheds light on the build-up of quasars at cosmic dawn'', INAF Minigrant 2024 ``The interstellar medium at high redshift'', and by the PRIN MUR ``2022935STW'', RFF M4.C2.1.1, CUP J53D23001570006 and C53D23000950006.
D.~Yang and J.~F.~Hennawi acknowledge support from the European Research Council (ERC) under the European Union’s Horizon 2020 research and innovation program (grant agreement No 885301).
J.~F.~Hennawi acknowledges support from NSF grant No. 2307180. 
M.O. is supported by the Japan Society for the Promotion of Science (JSPS) KAKENHI Grant Number 24K22894.
F.~Guarneri and J.-T.~Schindler are supported by the Deutsche Forschungsgemeinschaft (DFG,
German Research Foundation) - Project number 518006966. 
S.~E.~I.~Bosman is supported by the Deutsche Forschungsgemeinschaft (DFG) under Emmy Noether grant number BO 5771/1-1.
F.~Wang acknowledges support from NSF award AST-2513040.

This work made use of \texttt{NumPy} \citep{harris2020array}, \texttt{SciPy} \citep{virtanen2020}, \texttt{Astropy} \citep{astropy:2013,astropy:2018,astropy:2022}, \texttt{InterferoPy} \citep{interferopy}, \texttt{mapping} and \texttt{clic} within \textsf{GILDAS} \citep{gildas2013}, \texttt{CASA} \citep[][]{mcmullin2007}

\end{acknowledgements}

%
%

\bibliography{myref_total}
%

  


%

\label{LastPage}
\end{document}